\documentclass[fleqn,usenatbib]{mnras}

\usepackage{newtxtext,newtxmath}

\usepackage[T1]{fontenc}

\DeclareRobustCommand{\VAN}[3]{#2}
\let\VANthebibliography\thebibliography
\def\thebibliography{\DeclareRobustCommand{\VAN}[3]{##3}\VANthebibliography}

\usepackage{graphicx}	
\usepackage{amsmath}	
\usepackage{amssymb}	

\title[Coupling of ionised gas and stars]{The SAMI Survey: Evidence for dynamical coupling of ionised gas and young stellar populations}

\author[C. Foster et al.]{
Caroline Foster,$^{1,2,3}$\thanks{E-mail: c.foster@unsw.edu.au}, Sam Vaughan,$^{3,4,5,6}$ Amelia Fraser-McKelvie,$^{3,7}$ Sarah Brough,$^{1,3}$ Julia J. Bryant,$^{2,3,8}$ 
\newauthor
Scott M. Croom,$^{2,3}$ Francesco D'Eugenio,$^{9,10}$ Brent Groves,$^{7}$ Iraklis S. Konstantopoulos,$^{11}$ \'Angel R. L\'opez-
\newauthor 
S\'anchez,$^{3,4,5}$ Sree Oh,$^{3,12,13}$ Matt S. Owers,$^{3,4,5}$ Sarah M. Sweet,$^{3,14}$ Jesse van de Sande,$^{2,3}$ Emily Wisnioski,$^{3,12}$ 
\newauthor
Sukyoung K. Yi,$^{13}$ Henry R. M. Zovaro$^{3,12}$
\\
$^{1}$School of Physics, University of New South Wales, Sydney, NSW 2052, Australia\\
$^{2}$Sydney Institute for Astronomy, School of Physics, A28, The University of Sydney, NSW, 2006, Australia\\
$^{3}$ARC Centre of Excellence for All Sky Astrophysics in 3 Dimensions (ASTRO 3D)\\
$^{4}$Astronomy, Astrophysics and Astrophotonics Research Centre, Macquarie University, Sydney, NSW 2109, Australia\\
$^{5}$School of Mathematical and Physical Sciences, Macquarie University, NSW 2109, Australia\\
$^{6}$Centre for Astrophysics and Supercomputing, School of Science, Swinburne University of Technology, Hawthorn, VIC 3122, Australia\\
$^{7}$International Centre for Radio Astronomy Research, The University of Western Australia, 35 Stirling Hwy, 6009 Crawley, WA Australia\\
$^{8}$Astralis-USydney, School of Physics, University of Sydney, NSW 2006, Australia\\
$^{9}$Kavli Institute for Cosmology, University of Cambridge, Madingley Road, Cambridge, CB3 0HA, United Kingdom\\
$^{10}$Cavendish Laboratory - Astrophysics Group, University of Cambridge, 19 JJ Thomson Avenue, Cambridge, CB3 0HE, United Kingdom\\
$^{11}$Independent scholar Wellington, New Zealand\\
$^{12}$Research School of Astronomy and Astrophysics, Australian National University, Cotter Road, Weston Creek, ACT 2611, Australia\\
$^{13}$Department of Astronomy and Yonsei University Observatory, Yonsei University, Seoul 03722, Republic of Korea\\
$^{14}$School of Mathematics and Physics, University of Queensland, Brisbane, QLD 4072, Australia\\
}

\date{Accepted 02/2023.}

\pubyear{2023}

\begin{document}
\label{firstpage}
\pagerange{\pageref{firstpage}--\pageref{lastpage}}
\maketitle

\begin{abstract}
We explore local and global dynamical differences between the kinematics of ionised gas and stars in a sample of galaxies from Data Release 3 of the SAMI Galaxy Survey. We find better agreement between local (i.e., comparing on a spaxel-to-spaxel basis) velocities and dispersion of gas and stars in younger systems as with previous work on the asymmetric drift in galaxies, suggesting that the dynamics of stars and ionised gas are initially coupled. The intrinsic scatter around the velocity and dispersion relations increases with increasing stellar age and mass, suggesting that subsequent mechanisms such as internal processes, divergent star formation and assembly histories also play a role in setting and altering the dynamics of galaxies. The global (flux-weighted) dynamical support of older galaxies is hotter than in younger systems. We find that the ionised gas in galaxies is almost always dynamically colder than the stars with a steeper velocity gradient. In absolute terms, the local difference in velocity dispersion is more pronounced than the local difference in velocity, possibly reflecting inherent differences in the impact of turbulence, inflow and/or feedback on gas compared to stars. We suggest how these findings may be taken into account when comparing high and low redshift galaxy samples to infer dynamical evolution.
\end{abstract}

\begin{keywords}
galaxies: kinematics and dynamics -- galaxies: stellar content
\end{keywords}

\section{Introduction}

Galaxies are complex structures composed of both dark and baryonic matter. The interaction of baryons in galaxies is thought to be regulated by a number of physical processes (e.g., gravitational interactions/collisions, gas inflow, outflow, star formation, etc; see e.g., \citealt{Schaye15, Taylor15, Beckmann17, Conselice22}) and these processes may apply to or involve multiple distinct ``phases'' of the baryons (i.e., stars, ionised gas, dust, molecular gas, e.g., \citealt{Tumlinson17}). Studying galaxies through observations of multiple baryonic phases allows for a more holistic picture and understanding of the physical mechanisms that have shaped galaxies throughout cosmic time.

Stars within galaxies are a compendium of successive generations. Some stars form within a galaxy, while others are accreted from neighbouring or merged systems \citep[e.g.,][]{Ibata94,Bell08,DSouza18,Helmi18,Boecker20,Casanueva22,Remus22}. Over time, stellar orbits mix and evolve through interaction with external systems and secular star-to-star interactions. Enriched gas from previous stellar generations or externally accreted ``pristine'' gas can dissipate and form new stars usually by (re)forming a disc, thereby rejuvenating the system and changing its dynamics \citep[e.g.,][]{Dekel09,Wright21}.
Combined studies of ionised gas and stellar kinematics have shown that the gas is usually kinematically colder and faster rotating than the stars \citep[e.g.,][]{Pizzella04}.

Direct studies of the dynamical evolution of galaxies are limited by the relative brightness of the different phases. While the ionised gas can be detected out to higher redshifts ($z=0.7-2.5$, e.g., \citealt{Stott16,ForsterSchreiber18,Wisnioski19,Tiley21}), measuring spectra of the faint stellar continuum in galaxies is challenging/prohibitively observationally costly beyond $z \sim 1$ \citep[e.g.,][]{vanderWelvanderMarel08,Belli15,Mendel15,Belli17,Newman18b,Belli19,Mendel20}. Furthermore, the most massive galaxies in the present day universe are quiescent. Since massive star-forming galaxies at high redshifts are often the progenitors of quiescent local galaxies (e.g., \citealt{Guglielmo15}), understanding how coupled or otherwise the ionised gas is with the stars initially and over time is important for inferring the dynamical evolution of consistent galaxy populations across cosmic time \citep[also see][for a study comparing dynamical mass estimates based on ionised gas and stars]{Straatman22}. It is thus desirable to further explore the validity of comparing ionised gas dynamics at high redshift with local stellar dynamics.

Recent findings tentatively suggest that the dynamical properties of stars of different ages in nearby galaxies may mirror that of the ionised gas across cosmic time. Using Schwarzschild orbital modelling \citep[e.g.,][]{Schwarzschild79,vandenBosch08,Zhu20,Thater22} of the lenticular galaxy NGC~3115, \citet{Poci19} found that the stellar velocity dispersion as a function of population age steadily increases, mirroring that of the ionised gas in populations of galaxies across a broad range of redshifts \citep{Kassin12,Wisnioski15,Ubler19}. Older stars, like the ionised gas in high redshift galaxies, exhibit hotter dynamics (i.e., more turbulence or random motions) than their younger/local counterparts. This is surprising given the vastly different methods and samples used and does not imply the two are linked and much less causal. Indeed, this initially compelling picture has since been confounded by increased scatter when adding more systems into the local dynamically modelled sample \citep{Poci21}, suggesting that other explanations and evolutionary mechanisms need to be considered.

The ``asymmetric drift'' observed in the Milky Way \citep[e.g.,][]{Golubov13}, M31 \citep[e.g.,][]{Quirk19}, MaNGA \citep{Shetty20} and DiskMass \citep{Martinsson13} galaxies demonstrates that older stars tend to be on hotter orbits than young stars in present day systems. 
This is consistent with expectations from large hydrodynamical simulations \citep[e.g.,][]{Quirk20} and simulations of the Milky Way \citep{Bird13}. In the latter, the dynamics of Milky Way stars are shown to reflect the local conditions at the time of their formation. This tendency of older stars being on hotter orbits is consistent with the findings that older galaxies have larger intrinsic flattening (i.e. indicative of globally hotter orbits) than their younger counterparts \citep{vandeSande18}.

While the mirroring of stellar and gas dynamics across cosmic time may reflect the conditions of the interstellar medium (ISM) at that redshift it could also be a by-product of longer opportunities for internal processes of dynamical heating (e.g., stellar migration, radial mixing, etc). For example, \citet{Okalidis22} and \citet{Hayden20} suggest, using simulations and observations respectively, that stellar migration triggered by interactions with non-axisymmetric structures like bars explain aspects of the distribution of ages, metallicities and orbits of stars in the Solar neighborhood.

Older systems also have more time for externally triggered heating processes such as mergers and interactions to build up over time. Indeed, \citet{Minchev13} suggest that both stellar migration and interactions may play a role. \citet{Mackereth18} find that a significant episode of past interaction is required to produce the bimodal $\alpha$-element distribution observed in e.g. the Milky Way \citep{Hayden15} and other galaxies \citep[e.g.,][]{Scott21}.
In other words, a hybrid explanation where a mix of both initial conditions and latent dynamical heating may need to be invoked. Indeed, in eight Local Group galaxies, \citet{Leaman17} find that the stellar velocity dispersion of local galaxies is largely set by the conditions at the redshift of formation with departure from this being dependent on stellar mass.

There is a myriad of observational evidence supporting the idea that the dynamics of gas and stars are initially coupled. Integral field spectroscopic surveys in particular have made it clear that there are dynamical and chemical evidence for a co-evolution of stars and gas. Using the MaNGA galaxy survey, \citet{Shetty20} showed that the asymmetric drift exhibits more variations in stellar populations older than 1.5 Gyr compared to younger populations. In parallel, and also in MaNGA galaxies, \citet{Greener22} find evidence for chemical co-evolution of the ionised gas and stars. \citet{Barat19} compared rotational and central dispersion of gas and stars in the Sydney-AAO Multi Integral field (SAMI) Galaxy Survey and find a steeper slope for the $S0.5-\log_{10} M_*$ relationship in stars compared to gas kinematics. More recently, \citet{Oh22} showed that the velocity dispersion of the ionised gas tends to be lower than that of the stars in SAMI galaxies, though caution that beam smearing may be playing a role in artificially enhancing differences. As the precursor to stars, cold gas kinematics can also be compared to stellar kinematics. Indeed, \citet{Quirk19} find that the \ion{H}{i} and CO gas rotates slower when compared with the main sequence stars ($\sim7.5$~Gyr) in the Andromeda galaxy.

In this work, we use SAMI data to explore how the dynamics of the ionised gas and stars compare locally within galaxies. We explore whether measured local and global dynamical differences are consistent with current thinking on the origin of asymmetric drift and the role of ``latent'' dynamical heating and/or the ISM conditions at the time of formation.

The paper is divided as follows. \S\ref{sec:data} presents the data, while our sample selection can be found in \S\ref{sec:sample}. \S\ref{sec:analysis} details our data analysis and results. A discussion and our conclusions can be found in \S\ref{sec:discussion} and \S\ref{sec:summary}, respectively.

Throughout this paper, we assume a $\Lambda$CDM cosmology with  $\Omega_{\rm m}=0.3$, $\Omega_{\lambda}$~$=$~$0.7$ and $H_0=70$ km s$^{-1}$ Mpc$^{-1}$, and a \citet{Chabrier03} initial mass function.

\section{Data}\label{sec:data}

\begin{figure*}
\centering
\includegraphics[width=18.cm]{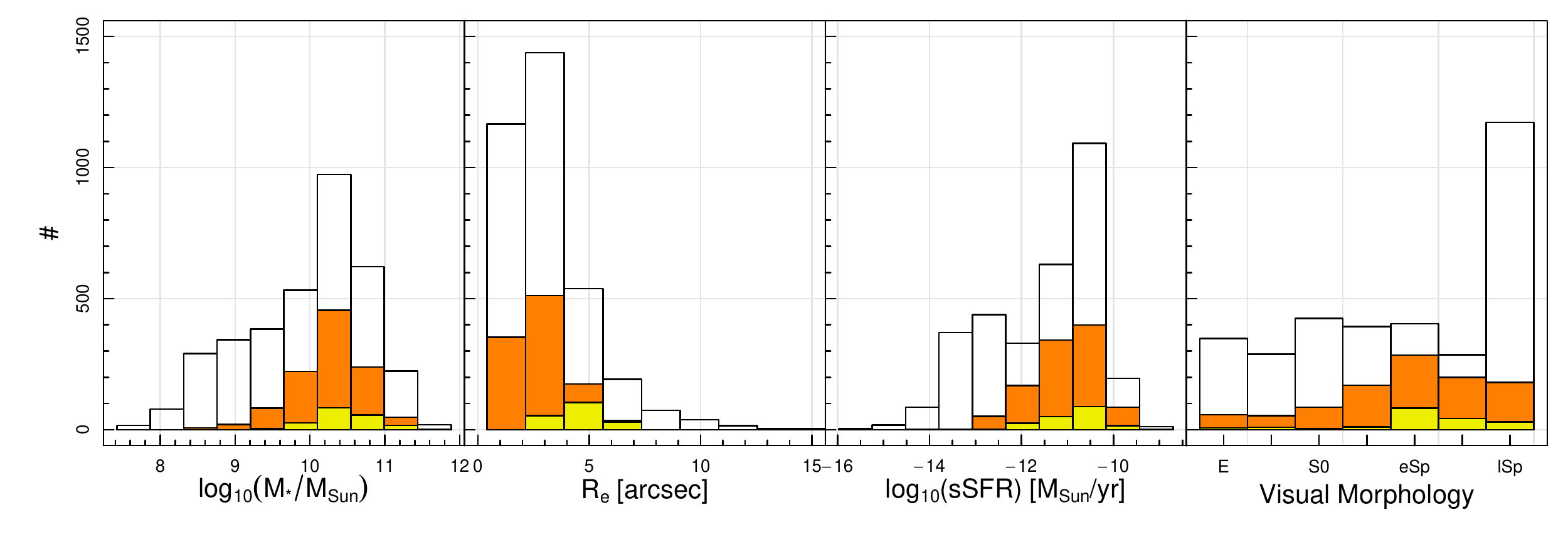}
\caption{Stellar mass ($\log_{10}(M_*/M_{\rm Sun})$), left), effective radius ($R_{\rm e}$, middle-left), specific star formation rate (sSFR, middle-right) and visual morphological classification (right, where E: elliptical, S0: lenticular, eSp: early spiral and lSp: late spiral) distributions for the final sample (yellow), galaxies with both gas and stellar kinematic maps out to 1 effective circularised radius but before the seeing and stellar mass cut are applied (orange) and the full SAMI sample (white). About $\sim 5$ percent of galaxies in full SAMI sample have uncertain morphological classifications and are not shown in the right panel. Due to the requirement of high quality, large spatial extent and good resolution of gas and stellar kinematic maps, the sample is necessarily biased towards star forming galaxies, intermediate morphological types (eSp) and higher effective radii.}
\label{fig:sample}
\end{figure*}

\begin{figure}
\centering
\includegraphics[width=8.5cm]{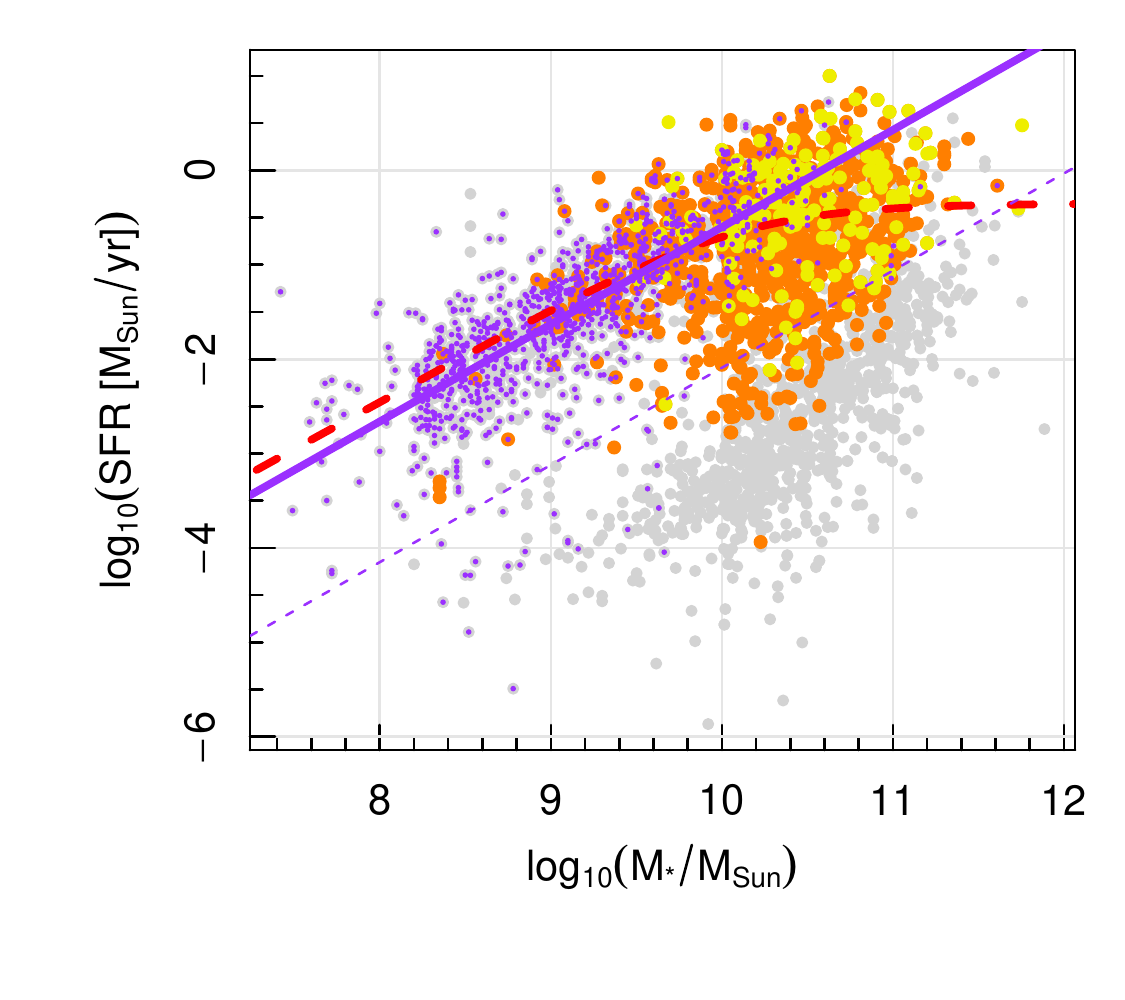}
\caption{Star formation rate as a function of the stellar mass for the final sample (yellow), galaxies with both gas and stellar kinematic maps out to 1 effective circularised radius but before the seeing and stellar mass cut are applied (orange) and the full SAMI sample (grey). The requirement for both gas and stellar kinematics favours galaxies with significant star formation, although there are a handful of selected galaxies that lie significantly ($>2$ standard deviations, dashed purple line) below the star forming main sequence defined by late-type spirals (purple dots) fitted as per \citet[][solid purple line]{Medling18}. Also shown as a red dashed line is the star forming main sequence as per Equation \ref{eq:SFMS}.}\label{fig:MSFR_sel}
\end{figure}

\defcitealias{Croom21}{C21}
This work uses spatially resolved spectroscopy and ancillary data from the Sydney-AAO Multi Integral field (SAMI) Galaxy Survey \citep[][henceforth \citetalias{Croom21}]{Croom21}. Spectral cubes were observed with SAMI \citep{Croom12}: a multi-object integral field instrument connected to the AAOmega spectrograph \citep{Sharp06} at the Anglo-Australian Telescope. SAMI had 13 hexabundles and 26 individual sky fibres deployable over a 1 degree field-of-view. Each hexabundle was a tightly packed bundle of 61 optical fibres over a 15 arcsec diameter with a 73 percent fill-factor. The SAMI Galaxy Survey observed $\sim 3000$ unique galaxies in the nearby Universe (i.e., redshifts $0.004< z <0.095$). The selection of the Galaxy And Mass Assembly (GAMA, \citealt{Driver11}) and clusters samples are detailed in \citet{Bryant15} and \citet{Owers17}, respectively. The data used in this work are part of the latest and final public SAMI Galaxy Survey data release (SAMI DR3, \citetalias{Croom21}). Thus, all data used in this work are available through this public data release.

SAMI data are reduced using the {\sc 2dfdr} pipeline, which performs the usual reduction steps: bias subtraction, wavelength calibration using CuAr arc frames and sky subtraction. Spectral extraction and reduction are detailed in \citet{Sharp10} and \citet{Hopkins13}, respectively. Flux calibration and telluric absorption corrections are performed using standard stars. Each field is nominally observed with a 7-dither (minimum 6) pattern to deal with gaps between fibres within the hexabundle. Data are then combined and mapped onto a grid with flux and covariance carefully propagated as described in \citet{Sharp15}. Further detail on the data reduction can be found in \citet{Allen15} and \citet{Sharp15}, with the latest improvements described in \citetalias{Croom21}.

Production of the stellar kinematic maps is described in \citet{vandeSande17a} with updates outlined in \citetalias{Croom21}. Briefly, the line-of-sight velocity distribution (LOSVD) is parametrised as a Gaussian using the penalised pixel-fitting ({\sc pPXF}; \citealt{Cappellari04}; \citealt{Cappellari17}) algorithm. A selection of templates are broadened and shifted through convolution to determine the most likely recession velocity ($V_{\rm rec}$) and velocity dispersion ($\sigma$) for each spaxel. This is done in two steps as described in \citet{vandeSande17a} to mitigate uncertainties associated with possible template mismatches, especially in the lower signal-to-noise spectra. The stellar kinematic position angles $PA_{\rm kin, stars}$ is determined using the {\sc fit\_kinematic\_pa} code, which is based on the method described in \citet[][their appendix C]{Krajnovic06}. 

Gas kinematics and H$\alpha$ emission line flux maps are measured using {\sc lzifu} \citep{Ho16} as described in \citet{Green18}, \citet{Scott18} and \citetalias{Croom21}. Briefly and as described in \citet{Owers19}, the continuum is first fitted to Voronoi-binned spaxels using single stellar population (SSP) templates from \citet{Vazdekis10} and \citet{GonzalezDelgado05}. This continuum is subtracted from individual spaxels to leave ``pure'' emission line spectra, which are then fitted using {\sc lzifu}.
Although {\sc lzifu} fits up to 3 independent velocity components, we only use the 1-component fit in this work. As for the stars, the kinematic position angle of the ionised gas ($PA_{\rm kin, gas}$) is determined using the {\sc fit\_kinematic\_pa} code following \citet[][their appendix C]{Krajnovic06}.

Global star formation rate (SFR) estimates are measured on elliptical $1R_e$ aperture spectra using the extinction-corrected H$\alpha$ flux and use the \citet{Kennicutt94} corrected to a \citet{Chabrier03} stellar initial mass function (\citealt{Medling18}; \citetalias{Croom21}, for more detail).



The light-weighted stellar population ages and metallicities ($[Z/H]$) within $1R_e$ are derived as described in \citet{Vaughan22}. We give a brief summary here. First, the SAMI blue and red arm spectra are joined together and convolved with a Gaussian kernel (of variable width) such that the spectral resolution is a constant value at all wavelengths (following \citealt{vandeSande17b}). We use \textsc{pPXF} \citep{Cappellari04, Cappellari17} to fit the MILES simple stellar population (SSP) models of \citep{Vazdekis15} to the spectrum of each galaxy extracted within $1R_e$. The templates range in metallicity from $-2.21$ dex to 0.4 dex, in age from 30 Myr to 14 Gyr and in [$\alpha$/Fe] abundance from 0.0 to $+0.4$ dex. The templates use the isochrones from the ‘Bag of Stellar Tracks and Isochrones’ models (BaSTI; \citealt{Pietrinferni04,Pietrinferni06}). 
During the fit, we also include gas emission line templates corresponding to the Balmer series (H$\alpha$, H$\beta$, and H$\gamma$) as well as the atomic species [\ion{N}{II}], [\ion{O}{III}], [\ion{S}{II}], and [\ion{O}{I}]. We use a multiplicative Legendre polynomial of the order of 10 to correct for small differences in the shape of the observed and template spectra. The final stellar population parameters for each spectrum are calculated following \citet{McDermid15}.

As described in \citet{dEugenio21} and \citetalias{Croom21}, photometric values such as effective radii,  position angles and ellipticities are obtained by applying the Multi-Gaussian Expansion technique \citep[MGE][]{Emsellem94} on $r$-band Sloan Digital Sky Survey and Very Large Telescope Survey Telescope images. 

Stellar masses are estimated following equation 8 of \citet{Taylor11} using $i$-band absolute magnitudes and $(g-i)$ colours. 

As described in \citep{Cortese16}, morphologies for all SAMI galaxies are based on careful compilation of visual classifications from multiple team members. A small fraction have uncertain visual morphological types ($\sim 5$ percent).

\section{Sample selection}\label{sec:sample}

Our target selection stringently keeps only those galaxies with the highest quality gas and stellar kinematic maps. Stellar kinematic map spaxels with $V_{\rm stars, error} > 30$ km s$^{-1}$, $\sigma_{\rm stars, error}>0.1\sigma+25$ km s$^{-1}$ and signal-to-noise ratio $<3$ are discarded. Gas kinematic maps spaxels with $F_{H\alpha}/F_{H\alpha, {\rm error}} < 5$ are discarded. We then only keep those galaxies where both the ionised gas and stellar kinematic maps have a minimum 85 percent fill factor within an elliptical aperture with circularised radius corresponding to 1$R_e$. As a result, all galaxies from the full SAMI sample of 3068 galaxies without significant and spatially extended emission lines are effectively removed. We select galaxies with $\log_{10} (M_{*}/M_{\rm Sun}) > 9.5$ and require that the full width at half maximum (FWHM) seeing value for the cube not exceed half the effective radius (i.e., $R_e > 2\times {\rm FWHM}$)\footnote{The median ${\rm FWHM}$ is 1.8 arcsec in our final sample.}.

The above selection leaves 188 unique galaxies and an additional 22 duplicate observations. For targets with duplicates, we select the instance with the better seeing. Fig. \ref{fig:sample} shows how the distributions in $\log_{10}(M_*/M_{\rm Sun})$, $R_e$, specific SFR (sSFR, i.e. the SFR per unit mass) and visual morphology of galaxies selected for this work compare with the full SAMI sample. Our requirement for high spatial resolution necessarily biases the sample towards high $R_e$ values and the requirement for reliable gas \emph{and} stellar kinematics biases the sample towards star forming and intermediate to later galaxy types (eSp). Fig. \ref{fig:MSFR_sel} shows the star formation rate (SFR) vs. stellar mass of SAMI galaxies. 
We note that the full SAMI sample is not a volume limited survey and that it includes many small and low-mass filler targets that do not pass our selection criteria.
The star-forming main sequence line is fit to the distribution of galaxies within the full SAMI sample with available SFRs as in \citet{FraserMcKelvie21}, using the functional form introduced by \citet{Leslie20}. Over the stellar mass range $8<\log_{10}(M_*/M_{\rm Sun})<11.5$, the best-fitting main sequence line was found to be:
\begin{equation}
\log_{10} {\rm SFR} = -0.352 - \log_{10}\left(1 + \frac{10^{10.101}}{M_*/M_{\rm Sun}}\right).
\end{equation}\label{eq:SFMS}
While the bulk of our selected galaxies lie along the star-forming main sequence, there are a considerable number that scatter to lower SFRs (Fig. \ref{fig:MSFR_sel}), consistent with at least some galaxies in our sample containing gas ionised by other ionising mechanisms than young massive hot stars.

\begin{figure}
\centering
\includegraphics[width=8.5cm]{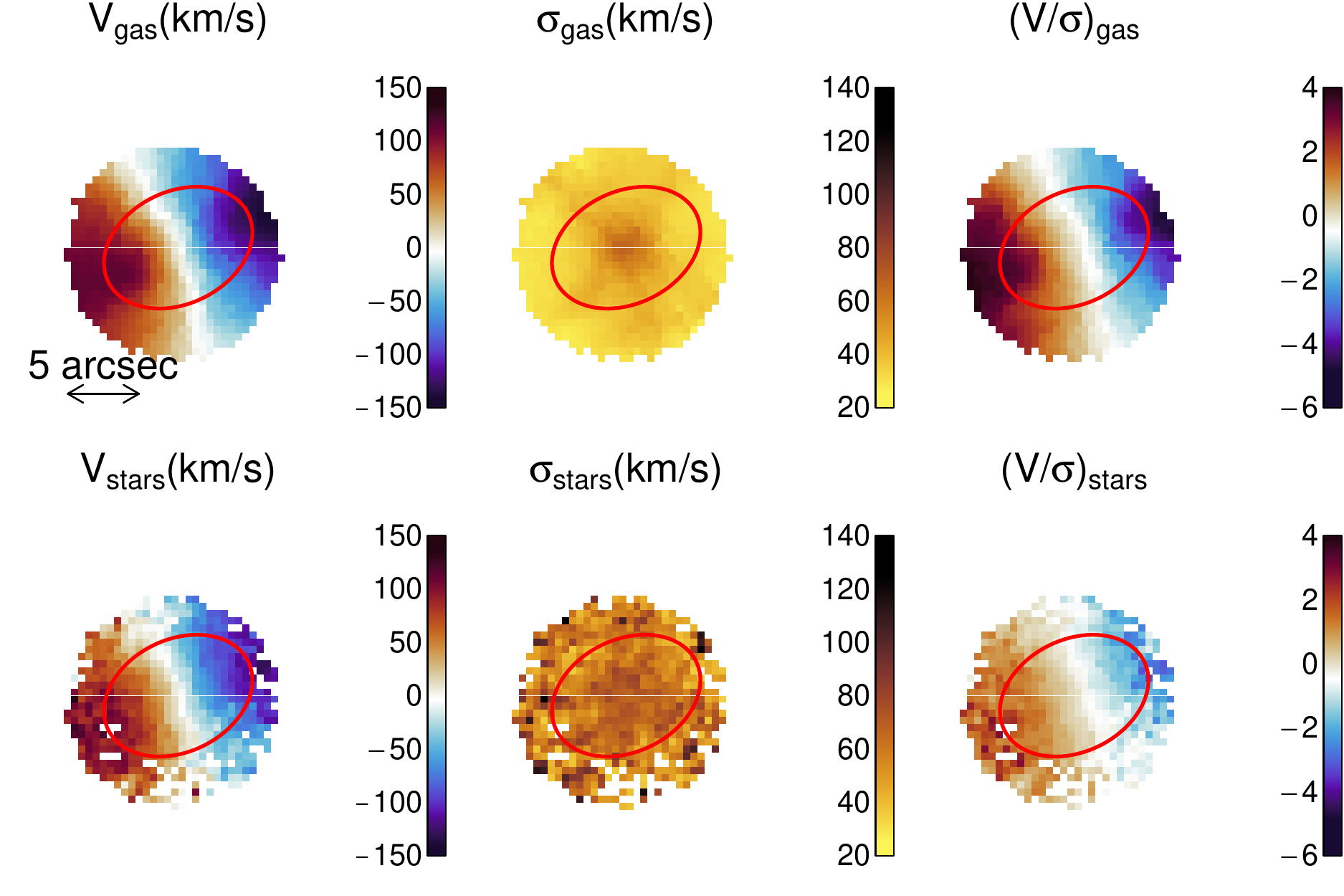}
\caption{Ionised gas (top) and stellar (bottom) velocity (left), dispersion (middle) and $(V/\sigma)$ maps for a ``typical'' galaxy in our final sample: SAMI106717. This target has $\Delta(V/\sigma)_{\rm R_e} = 0.88 \pm 0.08$, close to the median value for our final sample (i.e. $\Delta(V/\sigma)_{\rm R_e, median} = 0.82 \pm 0.04$). Red ellipses show the circularised effective radius aperture centered on the SAMI cube. In this example, the ionised gas exhibits higher rotational support than the stars. A scale is provided for reference on the top-left panel. North is up and East is left.}\label{fig:velfields}
\end{figure}

\section{Analysis and results}\label{sec:analysis}

\begin{table}
\centering
\begin{tabular}{ c|ccccc }
 \hline\hline
 Mean Age [Gyr] & 1 & 2 & 3 & 5 & 9 \\ 
 Mean $\left<\Delta V\right>_{R_e}$ [km s$^{-1}$] &  21 & 18 & 22 & 27 & 43\\
 std. dev. $\left<\Delta V\right>_{R_e}$ [km s$^{-1}$] & 11 & 6 & 11 & 29 & 34\\
 Mean $\left<\Delta \sigma\right>_{R_e}$ [km s$^{-1}$] & -35 & -37 & -40 & -47 & -53\\
 std. dev. $\left<\Delta \sigma\right>_{R_e}$ [km s$^{-1}$] & 12 & 12 & 18 & 20 & 30\\
 \hline\hline
Mean $\log_{10}(M_*/M_{\rm Sun})$ & 9.9 & 10.3 & 10.4 & 10.7 & 11.0\\
 Mean $\left<\Delta V\right>_{R_e}$ [km s$^{-1}$] &  23 & 18 & 26 &  28 & 39\\
 std. dev. $\left<\Delta V\right>_{R_e}$ [km s$^{-1}$] & 16 & 7 & 20 & 20 & 37\\
 Mean $\left<\Delta \sigma\right>_{R_e}$ [km s$^{-1}$] & -36 & -39 & -38 & -47 & -54\\
 std. dev. $\left<\Delta \sigma\right>_{R_e}$ [km s$^{-1}$] & 15 & 14 & 12 & 23 & 28\\
 \hline \hline
\end{tabular}
\caption{Age and stellar mass binned mean and standard deviation $\left<\Delta V\right>$ and $\left<\Delta \sigma\right>$ values as shown in Fig. \ref{fig:agesmstar_deltakin} \label{table:mean_deltas}}
\end{table}

\begin{figure*}
\centering
\includegraphics[width=9cm]{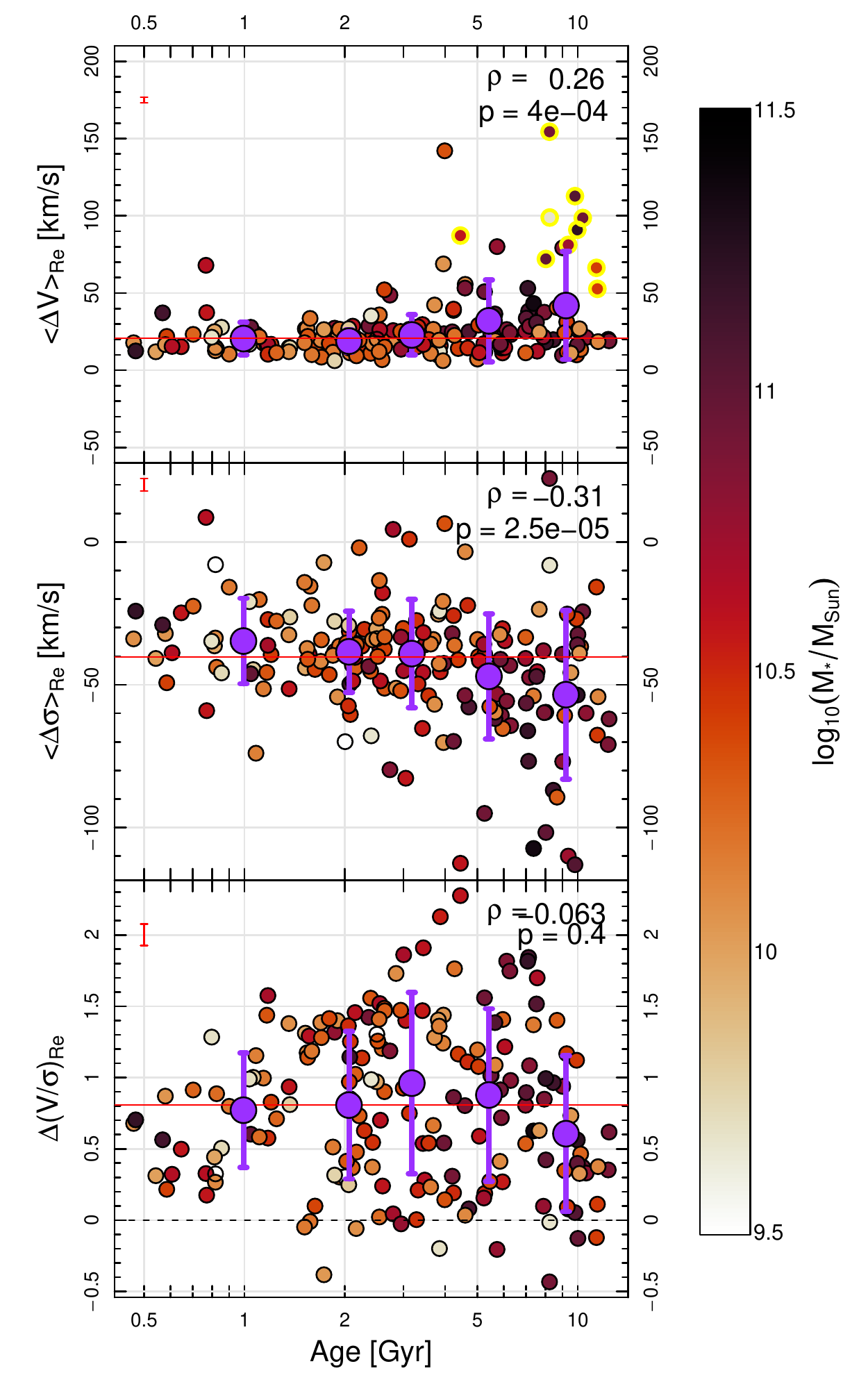}\includegraphics[width=9cm]{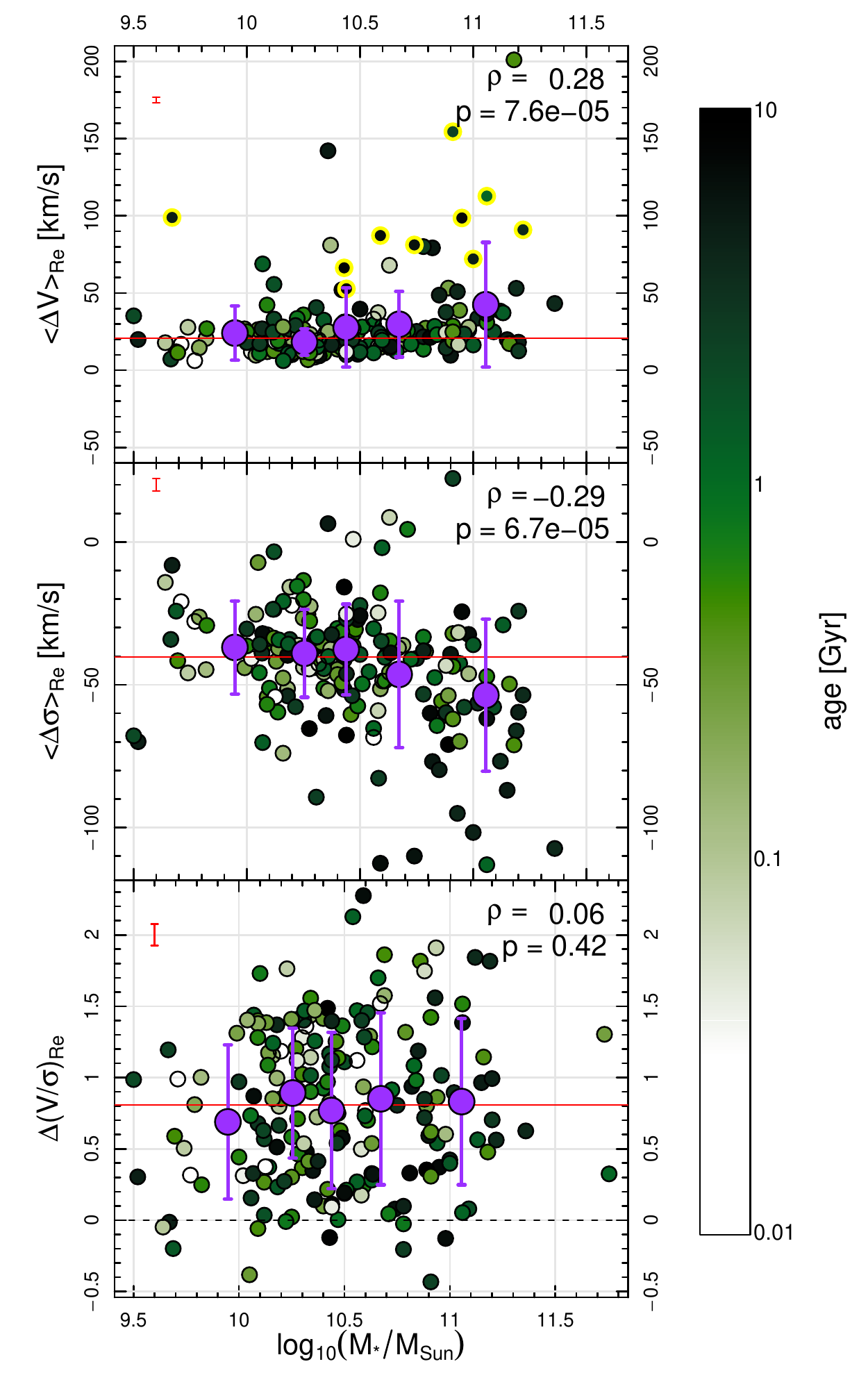}
\caption{Mean difference between the absolute gas and stellar velocity ($\left<\Delta V\right>_{R_e}$, top) and dispersion ($\left<\Delta\sigma\right>_{R_e}$, middle) as a function of stellar age (left) and stellar mass (right). The bottom panels show global $\Delta(V/\sigma)_{\rm R_e}$ as a function of stellar age (left) and stellar mass (right). Data points are colour-coded by stellar mass (left) and stellar age (right). Data points with yellow outlines in the top panel show systems with kinematically misaligned gas and stars. Red lines show median $y$-axis values in each panel. Median uncertainties on individual measurements are shown in red in the top left of each panel. Age-binned (left) and stellar mass-binned (right) mean values are shown as purple symbols with errorbars corresponding to the standard deviation of each bin (values recorded in Table \ref{table:mean_deltas}). The dashed lines represent $\Delta(V/\sigma)_{\rm R_e}=0$. The Spearman correlation coefficient $\rho$ and the corresponding $p$-value stated in the top right of each panel indicate a significant (anti-)correlation between age and $\left<\Delta V \right>_{R_e}$ ($\left<\Delta \sigma \right>_{R_e}$). There is increasing scatter in $\left<\Delta V \right>_{R_e}$ and $\left<\Delta \sigma \right>_{R_e}$ as a function of age. A significant (anti-)correlation between stellar mass and $\left<\Delta V \right>_{R_e}$ ($\left<\Delta \sigma \right>_{R_e}$) is observed. Galaxies with lower stellar masses have more similar gas and star dynamics. In the majority of systems, the ionised gas rotates faster than the stars (i.e., most data points and the median in the bottom panel lie above the dashed line). Interpretation is given in \S\ref{sec:discussion}.}\label{fig:agesmstar_deltakin}
\end{figure*}

Example gas and stellar velocity (V), dispersion ($\sigma$) and $(V/\sigma)$ maps for SAMI106717, a typical galaxy in our final sample, are shown in Fig. \ref{fig:velfields}. For this galaxy, and as we will show below, for most galaxies in our sample, the ionised gas is visibly dynamically colder than the stars (i.e., $(V/\sigma)_{\rm gas} > (V/\sigma)_{\rm stars}$).


We measure and subtract the systemic velocity ($V_{\rm sys}$) for the gas and stars separately by performing a flux weighted mean of the velocity map within 1.5 arcsec (3$\times$3 pixels).

To quantify the \textit{local} dynamical differences between the ionised gas and stars, we define two parameters. We purposely avoid the central regions to ensure that differences aren't driven by beam smearing, whose effect is most pronounced in the centre. First, we take the mean absolute velocity difference between gas and stars $\left<\Delta V\right>_{R_e}$ within an elliptical annulus of $0.9-1R_e$ around the SAMI cube centre, as:
\begin{equation}\label{eq:deltaV}
\left<\Delta V\right>_{R_e}=\frac{\sum_{0.9R_e\le R_i\le1R_e}{\left(\left| V_{{\rm gas}, i}-V_{{\rm stars},i}\right|\right)}}{\sum_{0.9R_e\le R_i\le1R_e}1},
\end{equation}
where $V_{{\rm gas},i}$ and $V_{{\rm stars},i}$ are the $V_{\rm sys}$-subtracted recession velocity measured in the $i^{\rm th}$ spaxel for the ionised gas and the stellar kinematic maps, respectively. $R_i$ is the circularised radius of the $i^{\rm th}$ pixel.

Similarly, the mean difference in velocity dispersion between the ionised gas and the stars, $\left<\Delta\sigma\right>_{R_e}$, is computed as follows:
\begin{equation}\label{eq:deltasigma}
\left<\Delta\sigma\right>_{R_e} = \frac{\sum_{0.9R_e\le R_i\le1R_e}{\left( \sigma_{{\rm gas}, i}-\sigma_{{\rm stars},i}\right)}}{\sum_{0.9R_e\le R_i\le1R_e}1},
\end{equation}
where $\sigma_{{\rm gas},i}$ and $\sigma_{{\rm stars},i}$ is the velocity dispersion measured in the $i^{\rm th}$ spaxel for the ionised gas and the stellar kinematic maps, respectively.

To study the \textit{global} nature (rotation vs pressure) of the dynamical support, we compute $(V/\sigma)$ for the ionised gas and stars within an elliptical $1R_e$ aperture following \citet{Cappellari07}:
\begin{equation}\label{eq:vos}
    (V/\sigma)_{X,R_e} = \sqrt{\frac{\left< V_{X,R_e}^2\right>}{\left< \sigma_{X,R_e}^2\right>}} = \sqrt{\frac{\sum_{R_i\le R_e}{F_i V_{X,i}^2}}{\sum_{R_i\le R_e}{F_i \sigma_{X,i}^2}}},
\end{equation}
where X stands for either `gas' or `stars'. $F_i$ is the mean continuum or H$\alpha$ flux for the stars or gas, respectively, in the $i^{\rm th}$ spaxel. We compute $\Delta(V/\sigma)$, the simple arithmetic difference between the ionised gas and stars $(V/\sigma)$, as follows:
\begin{equation}\label{eq:deltavos}
    \Delta(V/\sigma)_{R_e}=(V/\sigma)_{{\rm gas},R_e}-(V/\sigma)_{{\rm stars},R_e}.
\end{equation}

When relevant, we use the Spearman correlation coefficient ($\rho$) and associated $p$-value to determine the significance of trends. We consider values of $0.01\le p \le 0.05$ as weak trends, with $p < 0.01$ being the threshold to call a trend statistically significant.

Throughout, we refer to $\left<\Delta V \right>$ and $\left<\Delta\sigma\right>$ as measures of the local difference between the dynamics of the ionised gas and stars; and to $\Delta (V/\sigma)$ as a measure of global dynamical differences. In all cases, the uncertainties are propagated by neglecting covariance which is unavailable for the velocity and dispersion maps. We check the impact of ignoring covariance on our conclusions by repeating the analysis using every other spaxel. This effectively ensures that none of the spaxels used in the calculation are correlated (i.e., no covariance). This exercise shows only minimal absolute differences of order $\lesssim 1$ km s$^{-1}$ in the binned means and standard deviations of $\left<\Delta V\right>_{R_e}$ and $\left<\Delta\sigma\right>_{R_e}$ compared to those listed in Table \ref{table:mean_deltas} and thus ignoring covariance does not affect our conclusions.

The local and global kinematic differences within $1R_e$ with respect to age and stellar mass are shown in Fig. \ref{fig:agesmstar_deltakin}. The amplitude of local differences between the velocity maps ($\left<\Delta V\right>_{\rm R_e}$) are less scattered in younger stellar populations (see Table \ref{table:mean_deltas}).
Younger systems ($\sim 1$ Gyr) have a mean $\left<\Delta V\right>_{R_e} \sim 21$ km s$^{-1}$, while older ($\sim 9$ Gyr) systems have mean $\left< \Delta V\right>_{R_e} \sim 43$ km s$^{-1}$. The scatter away from this trend increases with stellar age with the standard deviation tripling from 11 km s$^{-1}$ for ages $\sim 1$ Gyr to 34 km s$^{-1}$ for ages $\sim 9$ Gyr.
Similarly, low mass systems ($\log_{10}(M_*/M_{\rm Sun}) \sim 9.9$) have a mean $\left<\Delta V\right>_{R_e} \sim 23$ km s$^{-1}$, while high mass ($\log_{10}(M_*/M_{\rm Sun}) \sim 11.0$) systems have mean $\left< \Delta V\right>_{R_e} = 39$ km s$^{-1}$. The scatter away from this trend more than doubles over the range of stellar masses probed (i.e. the standard deviation is 16 km s$^{-1}$ for $\log_{10}(M_*/M_{\rm Sun}) \sim 9.9$ and 37 km s$^{-1}$ for $\log_{10}(M_*/M_{\rm Sun}) \sim 11.0$).

\begin{figure}
\centering
\includegraphics[width=8.5cm]{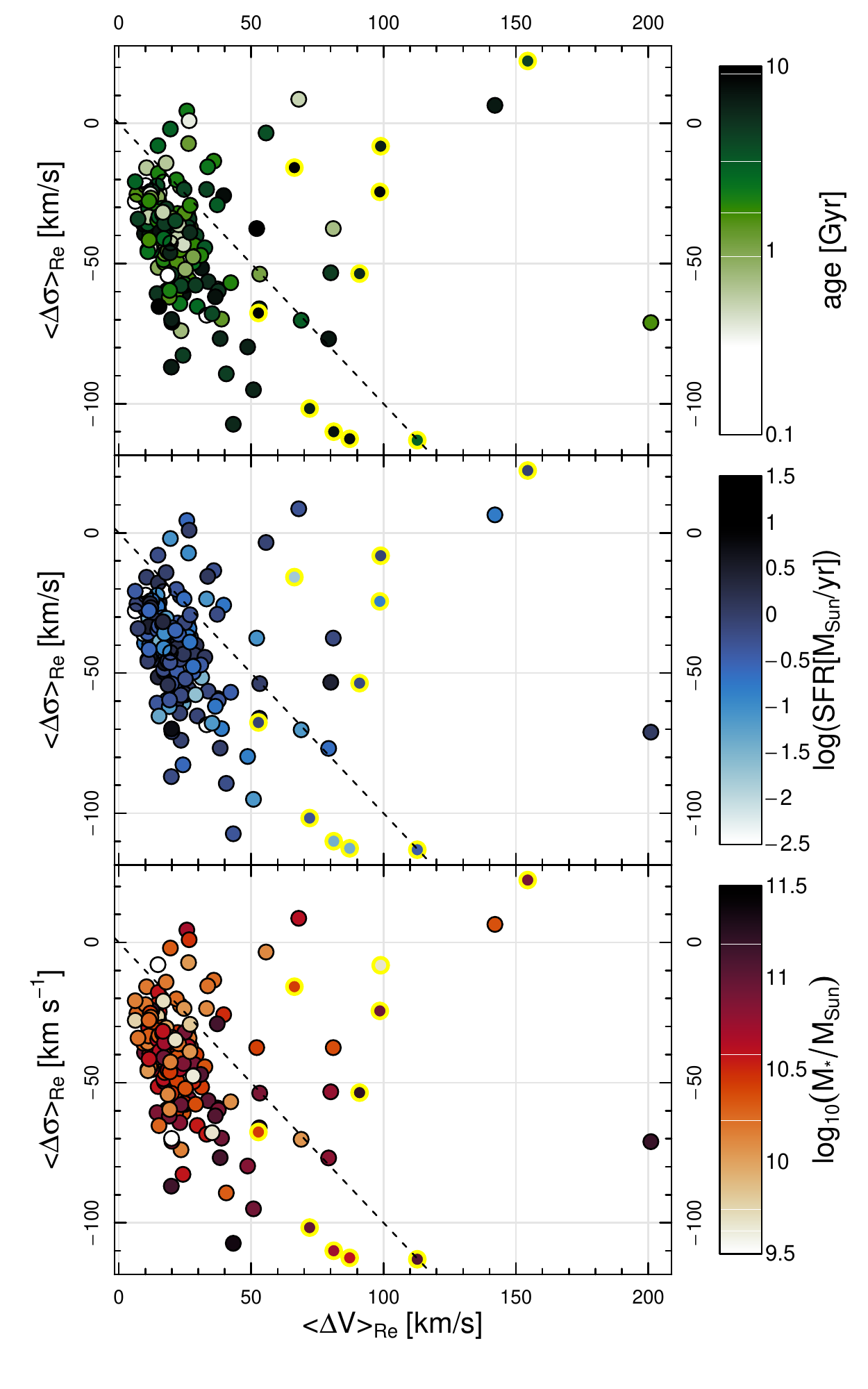}
\caption{The relationship between the mean difference between the ionised gas and stellar velocity ($\left< \Delta V \right>_{\rm R_e}$) and dispersion ($\left< \Delta \sigma \right>_{\rm R_e}$). Data points with yellow outlines show systems with kinematically misaligned gas and stars. The majority of data points lie below the line $\left< \Delta \sigma \right>_{\rm R_e}=-\left< \Delta V \right>_{\rm R_e}$ line (dashed). There is no clear trend with age or SFR. However, high mass galaxies tend to scatter lower in this space. Median values shown in Fig. \ref{fig:agesmstar_deltakin} indicate that in most systems, it is the localised dispersion, rather than the velocity, that differs most between the ionised gas and stars. Interpretation is given in \S\ref{sec:discussion}.}\label{fig:deltadelta}
\end{figure}

\begin{figure}
\centering
\includegraphics[width=9cm]{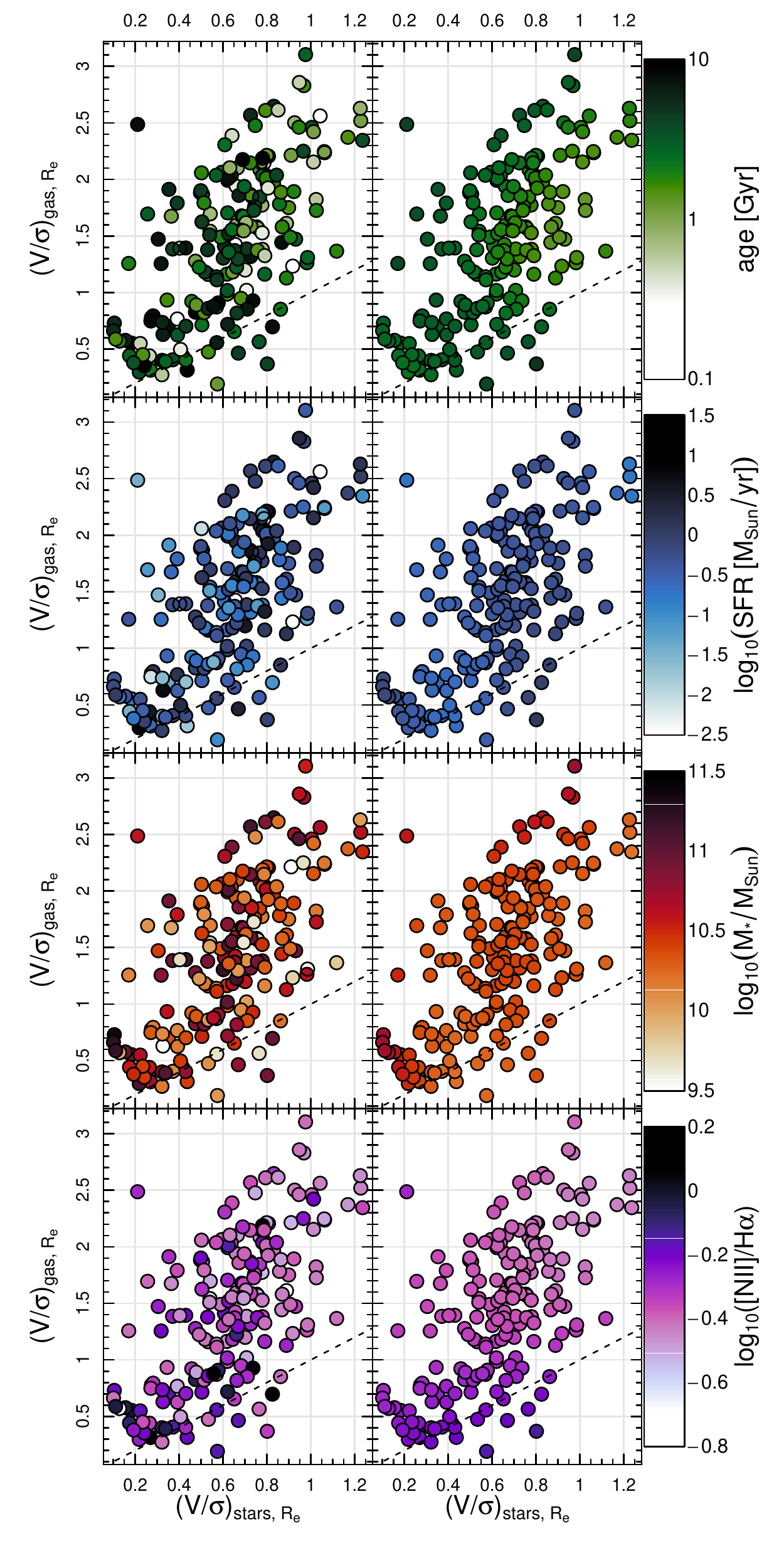}
\caption{LHS: Comparison of gas and stellar $(V/\sigma)_{R_e}$ with points colour-coded by stellar population age (top), SFR (second from top), stellar mass (second from bottom) and [\ion{N}{ii}$]/$H$\alpha$ (bottom). RHS: same as LHS, but using LOESS smoothing to highlight colour gradients. The ionised gas dynamics tend to be colder than the stars with a steeper slope than the one-to-one (dashed line). A colour gradient with lower values of $(V/\sigma)_{R_e}$ typically corresponding to older stellar ages and lower SFR on average is seen. A colour gradient suggests that higher values of [\ion{N}{ii}$]/$H$\alpha$ (consistent with LINER and/or AGN ionising radiation) correspond primarily to lower values of $(V/\sigma)_{\rm gas, R_e}$. No colour gradient is observed with stellar mass (third from top panels). On average, the global dynamics of ionised gas and stars better agree at lower $(V/\sigma)_{R_e}$ and for galaxies that are older, less star forming and with higher [\ion{N}{ii}$]/$H$\alpha$. Interpretation is given in \S\ref{sec:discussion}.}\label{fig:vosvsvos}
\end{figure}

The lower panels of Fig. \ref{fig:agesmstar_deltakin} indicate that the dynamics of the ionised gas is usually colder than the stars in the majority of systems (i.e. all but a handful of targets lie above the dashed line).
This is also shown in Fig. \ref{fig:vosvsvos}, which exhibits a steep correlation between $(V/\sigma)_{\rm stars, R_e}$ and $(V/\sigma)_{\rm gas, R_e}$, where the bulk of the data points lie above the one-to-one line. The trend is also steeper than the one-to-one, with older and/or lower SFR and/or higher [\ion{N}{ii}]/H$\alpha$ galaxies having lower and more similar $(V/\sigma)_{\rm gas, R_e}$ and $(V/\sigma)_{\rm stars, R_e}$.
The latter suggests a different mechanism (other than star-formation) is responsible for ionising the gas in those galaxies.

As per recent SAMI work by \citet{Ristea22} and similarly to other studies \citep[e.g.,][]{Davis11,Lagos15,Bryant19,Casanueva22}, we define misaligned galaxies as those where the kinematic position angle of the gas and stars differ by more than 30 degrees (i.e. $| PA_{\rm kin, gas}-PA_{\rm kin, stars}| > 30$). We find a total of 10 misaligned galaxies amongst the 188 galaxies in our final sample (namely, SAMI IDs 202399, 208652, 238922, 279905, 321059, 39057, 517278, 536994, 560238 and 618993).

Much of the increase in the scatter in $\left<\Delta V\right>_{R_e}$ as a function of mass and age corresponds to a higher proportion of kinematically misaligned galaxies (yellow outlined symbols in Fig. \ref{fig:agesmstar_deltakin}). In other words, higher scatter tends to correspond to more massive and older galaxies, with all misaligned galaxies having flux weighted stellar ages $>3$ Gyr.



Fig. \ref{fig:deltadelta} shows how the local dispersion differs more than the local velocity in most galaxies (except for kinematically misaligned systems) and that this does not seem to be linked with stellar population age or SFR, but possibly weakly with stellar mass.

\begin{figure}
\centering
\includegraphics[width=8.5cm]{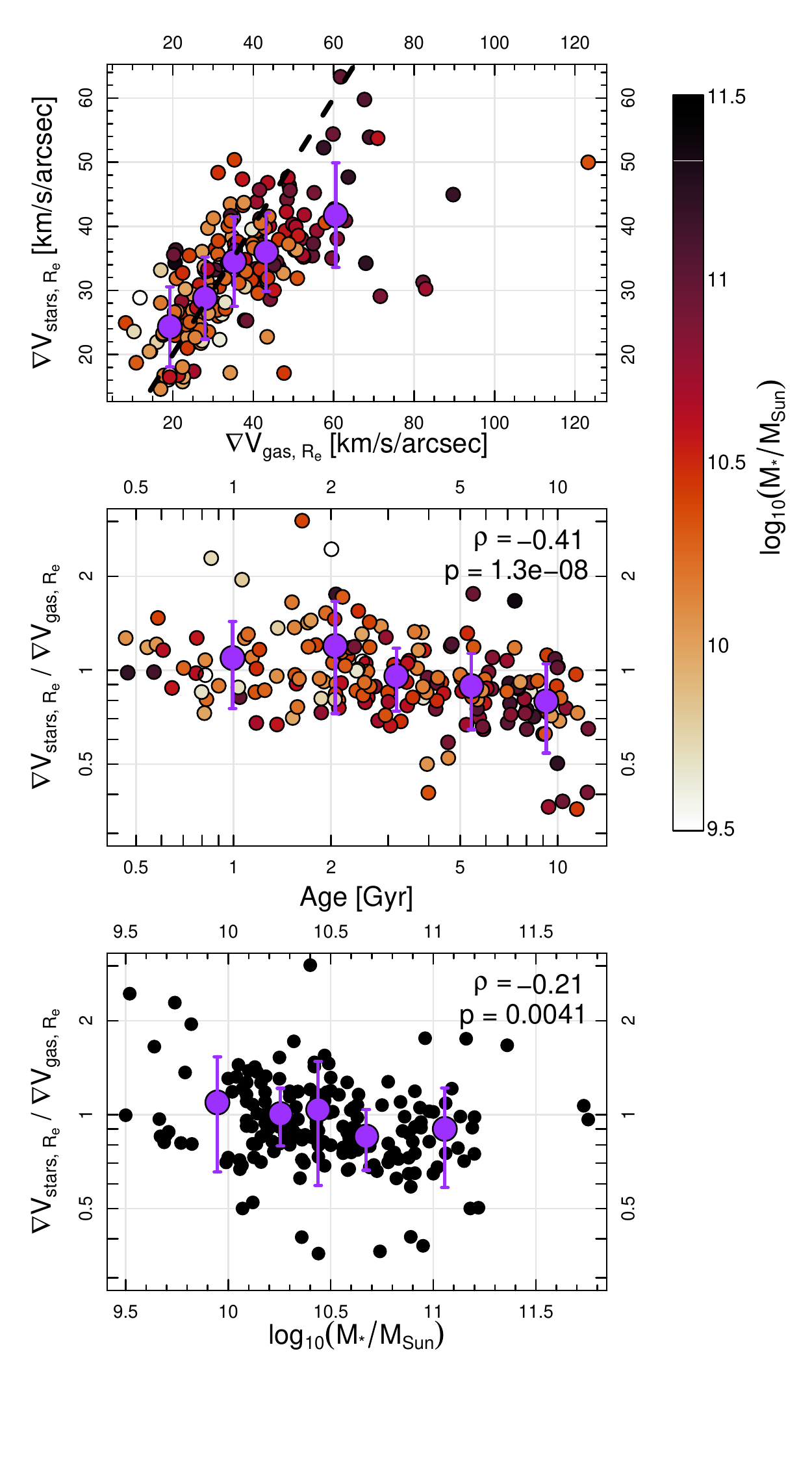}
\caption{Top: comparison of the gas ($\nabla V_{\rm gas, R_e}$) and stellar ($\nabla V_{\rm stars, R_e}$) velocity gradients. Dashed line is the one-to-one. Middle: ratio of the stars and gas velocity gradients as a function of stellar age. Bottom: same as middle, but as a function of stellar mass. In the upper two panels, symbols are colour coded by stellar mass. Large purple symbols represent mean values binned along the x-axis with errorbars representing the standard deviation in each bin. The Spearman correlation coefficient $\rho$ and the corresponding $p$-value stated in the top right of the bottom two panels indicate significant anti-correlations between stellar mass and the ratio of the velocity gradients. Velocity gradients best agree in low mass and younger systems. In high mass and older systems, the gas velocity gradient is on average steeper than the stellar velocity gradient. Interpretation is given in \S\ref{sec:discussion}.} \label{fig:age_gradongrad}
\end{figure}

Following \citet{Varidel16} and \citet{Oh22}, we compute the projected mean velocity gradient within an elliptical $1R_e$ aperture for both the ionised gas and stellar kinematics as follows:
\begin{equation}
    \nabla V_{\rm X, R_e} = \frac{1}{2W_{\rm pix}} \sqrt{\frac{\sum_{R_i \le R_e}{ \left(\begin{aligned}& [V(x_i+1,y_i)-V(x_i-1,y_i)]^2 \\ & + [V(x_i,y_i+1) - V(x_i, y_i-1)]^2 \end{aligned}\right)}
    }{N_{\rm pix, R_e}}},
\end{equation}
where $W_{\rm pix}$ is the pixel scale (i.e., 0.5 arcsec/pix for SAMI) and $N_{\rm pix, R_e}$ is the number of pixels within $1R_e$. $\nabla V_{\rm gas, R_e}$ and $\nabla V_{\rm stars, R_e}$ reflect the spatial change within the effective radius of the velocity of galaxies. The velocity gradient has been used as a proxy for the impact of beam smearing on gas and stellar velocity dispersion. In Appendix \ref{sec:beamsmearing}, we discuss how our results are robust against the impact of beam smearing and that trends with $\nabla V$ at least partly reflect real physical processes.

Fig. \ref{fig:age_gradongrad} shows that the projected gas and stars velocity gradients best agree in low mass and younger systems. Conversely, in high mass and older systems, the gas velocity gradient is on average steeper than the stellar velocity gradient.

\section{Discussion}\label{sec:discussion}

The main driver for this work is to better understand how gas and stars in galaxies correlate, and test whether one may safely assume that the two phases are coupled, at least initially, when forming stars. In other words, can we infer dynamical evolution of galaxies using high and low redshift surveys by comparing the dynamics of ionised gas in massive star forming galaxies to their local often quiescent (and without ionised gas) descendants? In what follows, we explore how this may be done along with stating significant caveats to this approach.

Despite our simplistic approach, our results are relevant to a number of topics in the literature, including asymmetric drift, dynamical heating, evolution of ISM dynamics, as well as our main goal, which is to test possible methodologies for quantifying the dynamical evolution of galaxies. As such, our results are relevant to both local group and high redshift studies. This discussion focuses on contrasting a selection of relevant previous work from the high and low redshift literature as well as local group studies to contextualise our findings.

As a first caveat, in galaxies with a complex mix of stellar populations, the stellar dynamics do not just represent those of the youngest stars. Hence, we do not necessarily expect an exact one-to-one correspondence between the ionised gas and stars in any of our systems. This is because on the one hand, newborn stars ionise and heat up the gas. On the other hand, the physics of gas and stars fundamentally differ, with gas being collisional and stars being collisionless, which affects the heating mechanisms of the two dynamical tracers. Furthermore, the ionising source of the gas can vary between systems and even spatially across a single galaxy. Ionising sources can include hot young and/or evolved stars, low-ionization nuclear emission-line region (LINER), shocks and Active Galactic Nuclei (AGN).

Using simulations of the Milky Way, \citet{Bird13} show that the dynamics of the stars could reflect the ISM conditions of their surroundings at the time of their formation.
Assuming massive galaxies are more dynamically ``evolved'', the trends we see in the mean $\left< \Delta V \right>_{R_e}$ and $\left< \Delta \sigma \right>_{R_e}$ values (i.e., large purple symbols in Fig. \ref{fig:agesmstar_deltakin}) are consistent with a scenario in which gas and stars are initially partly coupled or in which heating mechanisms follow an overall trend across cosmic time. In other words, in young and low mass galaxies, the dynamics of the gas and stars could be more similar if the gas from which stars form is being ionised by (possibly those same) recently formed stars with possible subsequent heating enhancing differences over time. This suggests that the stars may continue to carry some of the kinematic information of the gas from which they formed and subsequently ionised. It is noteworthy that \citet[][for a single galaxy]{Poci19} found an offset of about $\sim10$ km s$^{-1}$ in the stellar dispersion compared to the ionised gas data from \citet{Wisnioski15} at the same epoch, while \citet{Poci21} found significant scatter in this offset when comparing with different systems.

In almost all cases, galaxies show colder gas dynamics than the stars, with the difference being more pronounced in high mass and older systems.
Our finding that galaxies with younger stellar populations (or higher star formation rate, see Fig. \ref{fig:vosvsvos}) have more pronounced differences in dynamical support, with a steeper than one-to-one relationship between dynamical support of the ionised gas and stars, is consistent with a scenario wherein galaxy evolution becomes increasingly driven by dry mergers just as the global star formation drastically reduces \citep[e.g.,][]{MadauDickinson14}, thereby heating the stars through orbital mixing.

The fact that the most massive and older galaxies tend to scatter more than their lower mass and younger counterparts (top panel of  Fig. \ref{fig:agesmstar_deltakin} and bottom panel of Fig. \ref{fig:deltadelta}) is consistent with the findings of \citet{Leaman17} and with scenarios in which stars experience varying degrees of dynamical ``latent'' heating (i.e., dynamical heating after formation) depending on the galaxies' individual merging and formation histories. This interpretation that dynamical heating could be the cause of the increased scatter is corroborated by the fact that the most highly scattered points tend to be those with misaligned gas and stars kinematics, which is primarily associated with external gas accretion processes as recently argued by e.g. \citet{Ristea22}.

An important caveat and consideration is that we use luminosity-weighted ages of mixed stellar populations, which are inherently uncertain and may over-represent the brighter younger populations as the mass-to-light ratio plateaus at older ages (e.g., \citealt{Bruzual03}), hence leading to a higher scatter in age. 

It may be possible to compare the local dynamical differences between gas and various stellar populations using orbital dynamical modelling for a subset of the galaxies in our sample (e.g., \citealt{Poci19}; \citealt{Santucci22}) and test the impact of using light-weighted ages on our conclusions, however this is beyond the scope of this work.

Comparing the dynamics of the ionised gas and stars locally (i.e., on a spaxel-by-spaxel basis) ensures that radial differences in asymmetric drift are implicitly accounted for \citep{Shetty20}. The increased scatter in $\left<\Delta V\right>_{\rm R_e}$ at older ages is consistent with the MaNGA results of \citet{Shetty20} that the asymmetric drift of stellar populations beyond 1.5~Gyr varies greatly, while the asymmetric drift of stellar populations younger than 1.5~Gyr is fairly stable with radius at or above 0 km s$^{-1}$. This is again consistent with and likely reflects a broad range of possible galaxy formation and evolutionary pathways that diverge slowly over time.


The measured local offsets (median$\left<\Delta V\right>_{\rm R_e}=+20\pm1$  km s$^{-1}$ and median $\left<\Delta\sigma\right>_{\rm R_e}=-40\pm1$ km s$^{-1}$) are of comparable order of magnitude to those found in \citet{Quirk19} for different gas phases in Andromeda. Namely, \citet{Quirk19} find that the \ion{H}{i} and CO gas rotates faster by up to 18 km s$^{-1}$ than the younger asymptotic giant branch stars ($\sim2$~Gyr). The difference increases to $63$ and 37 km s$^{-1}$ for \ion{H}{i} and CO gas, respectively, when compared with red giant branch stars ($\sim4$~Gyr).

The finding that the dispersion of the ionised gas and stars (median $\left<\Delta\sigma\right>_{\rm R_e}$ is $-40$ km s$^{-1}$) differs more (in absolute terms) than the velocities (median $\left<\Delta V\right>_{\rm R_e}$ is $+20$  km s$^{-1}$; see Figs. \ref{fig:agesmstar_deltakin} and \ref{fig:deltadelta}) may be a result of inherent differences in the collisional properties of gas and stars and the impact of inflow and/or non star-forming feedback activity mainly on gas dynamics/heating (as described in e.g., \citealt{Wellons20}). Part of this difference may also be explained by beam smearing effects (see Appendix \ref{sec:beamsmearing}).

Also using SAMI data, \citet{Oh22} found that gas emission not directly associated with star-formation (e.g., AGN/LINER) is associated with higher gas dispersion. This finding could explain why, in galaxies with older stellar populations (i.e., those more likely to harbour AGN or LINER-like emission at the expense of star formation), the measured $(V/\sigma)_{\rm gas}$ is lower (see lower panel of Fig. \ref{fig:agesmstar_deltakin}). If the reason that the ionised gas dynamics are hotter in galaxies with older stellar populations than younger ones is because the gas is, on average, ionised by different sources, they should also exhibit higher [\ion{N}{ii}$]/$H$\alpha$, which is confirmed in the bottom panel of Fig. \ref{fig:vosvsvos}. The higher gas dispersion of the non star-forming gas could be associated with outflows in the case of AGN and shocks, but could also be associated with the diffuse ionised gas, that shows LINER-like emission and has both a higher dispersion and slower velocity structure when extra-planar  \citep[see eg, ][]{Levy19, Belfiore22,Micheva22}.

Another important dimension of this work is the observed difference in the velocity gradients ($\nabla V$) of the ionised gas and stars. 
Our results indicate that in most galaxies, the stellar velocity gradient is steeper than the gas velocity gradient in those systems. This is true at nearly all masses, except in high mass galaxies where it is the ionised gas gradient that is steeper. An important caveat of using $\nabla V$ as defined here is that the ``physical'' pixel scale (i.e. kpc per pixel) changes with redshift and hence across the sample and this is not directly taken into account. While $\nabla V_{\rm gas}$ has been used as a beam smearing proxy \citep{Oh22,Varidel16}, Fig. \ref{fig:age_gradongrad} shows that there are statistically significant differences in the gradient of gas and stars that also reflect differences in the dynamics of gas and stars with stellar age and stellar mass discussed previously.

In Appendix \ref{sec:beamsmearing}, we investigate the possible impact of beam smearing on our results using two separate proxies. \citet{Oh22} suggested that gas velocity dispersion in SAMI is more impacted by beam smearing than stellar dispersion owing to steep gas velocity gradients using a different sample selection and $\Delta\sigma$ parameter definition. In the appendix, we argue that our stringent selection for the best resolved targets and different parameter definitions mitigate the impact of beam smearing and discuss that some of the trends seen with various beam smearing proxies are at least partly attributed to the real physical effects discussed.

As a sanity check, we also investigate possible trends with the uncertainties in the kinematics scale with age or stellar mass (not shown). As expected, we do find a shallow trend with stellar mass, but this trend does not explain the magnitude of the scatter shown in e.g., Fig. \ref{fig:agesmstar_deltakin}, and there is no trend of uncertainty with age.

\section{Summary and conclusions}\label{sec:summary}

In this work, we explore the difference in local and global dynamics of the ionised gas and stars. We select a sample of 188 SAMI galaxies with high quality kinematic maps and optimal spatial resolution. Local differences ($\left<\Delta V\right>$ and $\left<\Delta\sigma\right>$) are computed for a $1R_e$ elliptical annulus on a spaxel-by-spaxel basis, while global differences ($\Delta (V/\sigma)$) are computed using \emph{flux weighting} within a $1R_e$ aperture.

We find that our findings are broadly consistent with observational and theoretical expectations from asymmetric drift (e.g., \citealt{Quirk20}) and detailed modelling of the stellar dynamics and populations in local galaxies (e.g., \citealt{Poci19,Poci21}).

Our main conclusions are summarised as follows:
\begin{itemize}
    \item The local dynamics of the ionised gas better mimic those of the stars in galaxies with younger light-weighted stellar population ages or lower mass than in older and higher mass systems. This is consistent with the dynamics of ionised gas and stars being initially coupled (Figs. \ref{fig:agesmstar_deltakin}, \ref{fig:deltadelta}).
    \item The two- to three-fold increase in scatter in local dynamical differences between gas and stars and the increased divergence of the gas and stellar velocity gradients with stellar mass and age are consistent with dynamical heating playing a key role and resulting from a broad range of formation and assembly histories among galaxies (Figs. \ref{fig:agesmstar_deltakin}, \ref{fig:age_gradongrad}).
    \item The global dynamics of the ionised gas are typically colder than those of the stars in the star-formation dominated galaxies. In galaxies with AGN/LINER-like emission, the global dynamics are more similar (Figs. \ref{fig:agesmstar_deltakin}, \ref{fig:vosvsvos}).
    \item Older galaxies have hotter stellar and ionised gas dynamics on average than younger galaxies, and we argue that this difference is consistent with being driven by different merging and/or accretion histories and/or ionising mechanisms (Fig. \ref{fig:vosvsvos}).
    \item In absolute terms, the median local difference in the velocity dispersion of the ionised gas and stars is greater than the difference in velocity regardless of stellar age (i.e., $|\left<\Delta\sigma\right>_{\rm R_e, median}| - |\left<\Delta V\right>_{\rm R_e, median}| = 20$ km s$^{-1}$; Fig. \ref{fig:deltadelta}). This may reflect inherent differences in the collisional properties of gas and stars or the impact of turbulence, inflow and/or non star-forming feedback on gas dynamics.
\end{itemize}

More generally, while younger stellar populations may initially better reflect the dynamical properties of the gas from which they formed, our findings suggest there would be inherent risks associated with comparing ionised gas kinematics at high redshifts with the stellar dynamics of local systems to infer the dynamical evolution of (now) quiescent galaxies. Considering sufficiently large samples and limiting the investigation to star-forming ionised gas can help mitigate the inherent scatter and counfounding factors identified in this work.

\section*{Acknowledgements}

We thank S. L. Martell for helpful comments and discussions.

CF is the recipient of an Australian Research Council Future Fellowship (project number FT210100168) funded by the Australian Government. M.S.O. acknowledges the funding support from the Australian Research Council through a Future Fellowship (FT140100255). S.K.Y. acknowledges support from the Korean National Research Foundation (2020R1A2C3003769, 2022R1A6A1A03053472). JvdS acknowledges support of an Australian Research Council Discovery Early Career Research Award (project number DE200100461) funded by the Australian Government. SMS acknowledges funding from the Australian Research Council (DE220100003). SB acknowledges funding support from the Australian Research Council through a Future Fellowship (FT140101166). JJB acknowledges support of an Australian Research Council Future Fellowship (FT180100231). FDE acknowledges funding through the ERC Advanced grant 695671 “QUENCH”, the H2020 ERC Consolidator Grant 683184 and support by the Science and Technology Facilities Council (STFC).

Part of this research was conducted by the Australian Research Council Centre of Excellence for All Sky Astrophysics in 3 Dimensions (ASTRO 3D), through project number CE170100013.

The SAMI Galaxy Survey is based on observations made at the
Anglo-Australian Telescope. The Sydney-AAO Multi-object Integral field spectrograph (SAMI) was developed jointly by the University of Sydney and the Australian Astronomical Observatory. The SAMI input catalogue is based on data taken from the Sloan Digital Sky Survey, the GAMA Survey and the VST ATLAS Survey. The SAMI Galaxy Survey website is \url{http://sami-survey.org/}. The SAMI Galaxy Survey is supported by the Australian Research Council Centre of Excellence for All Sky Astrophysics in 3 Dimensions (ASTRO 3D), through project number CE170100013, the Australian Research Council Centre of Excellence for All-sky Astrophysics (CAASTRO), through project number CE110001020, and other participating institutions.

Based on data acquired at the Anglo-Australian Telescope under programs A/2013B/012 and A/2016B/16. We acknowledge the traditional owners of the land on which the AAT stands, the Gamilaraay people, and pay our respects to elders past and present.

GAMA is a joint European-Australasian project based around a
spectroscopic campaign using the Anglo-Australian Telescope. The GAMA
input catalogue is based on data taken from the Sloan Digital Sky
Survey and the UKIRT Infrared Deep Sky Survey. Complementary imaging
of the GAMA regions is being obtained by a number of independent
survey programmes including GALEX MIS, VST KiDS, VISTA VIKING, WISE,
Herschel-ATLAS, GMRT and ASKAP providing UV to radio coverage. GAMA is
funded by the STFC (UK), the ARC (Australia), the AAO, and the
participating institutions. The GAMA website is
\url{http://www.gama-survey.org/}.

This work makes use of colour scales chosen from \citet{CMASHER2020}. 

\section*{Data Availability}

All SAMI DR3 data \citep{Croom21} used in this work are publicly available through Data Central (\url{datacentral.org.au}). SAMI DR3 data tables used in this work are: {\sc EmissionLine1CompDR3}, {\sc MGEPhotomUnregDR3}, {\sc InputCatGAMADR3}, {\sc InputCatClustersDR3} and {\sc VisualMorphologyDR3}. The H$\alpha$ flux and signal-to-noise maps are based on ``SAMI DR3 1-component line emission map: H$\alpha$''. The [\ion{N}{ii}] flux maps are taken from ``SAMI DR3 1-component line emission map: [NII](6583\AA)''. Kinematic maps used for the ionised gas are the ``SAMI DR3 1-component ionised gas velocity map'' and ``SAMI DR3 1-component ionised gas dispersion map''. Stellar kinematic maps used are ``SAMI DR3 Stellar Velocity map (2 moments) default cube'' and ``SAMI DR3 Stellar Velocity dispersion map (2 moments) default cube''. Stellar ages were contributed by co-author S. Vaughan based on the publicly available SAMI DR3 spectra as described in \S\ref{sec:data}.



\bibliographystyle{mnras}
\bibliography{biblio} 



\appendix

\section{Beam smearing}\label{sec:beamsmearing}

In this section, we explore the possible impact of beam smearing (or atmospheric seeing) on our results and conclusions. We investigate two separate proxies for the impact of beam smearing.

\begin{figure}
\centering
\includegraphics[width=8.5cm]{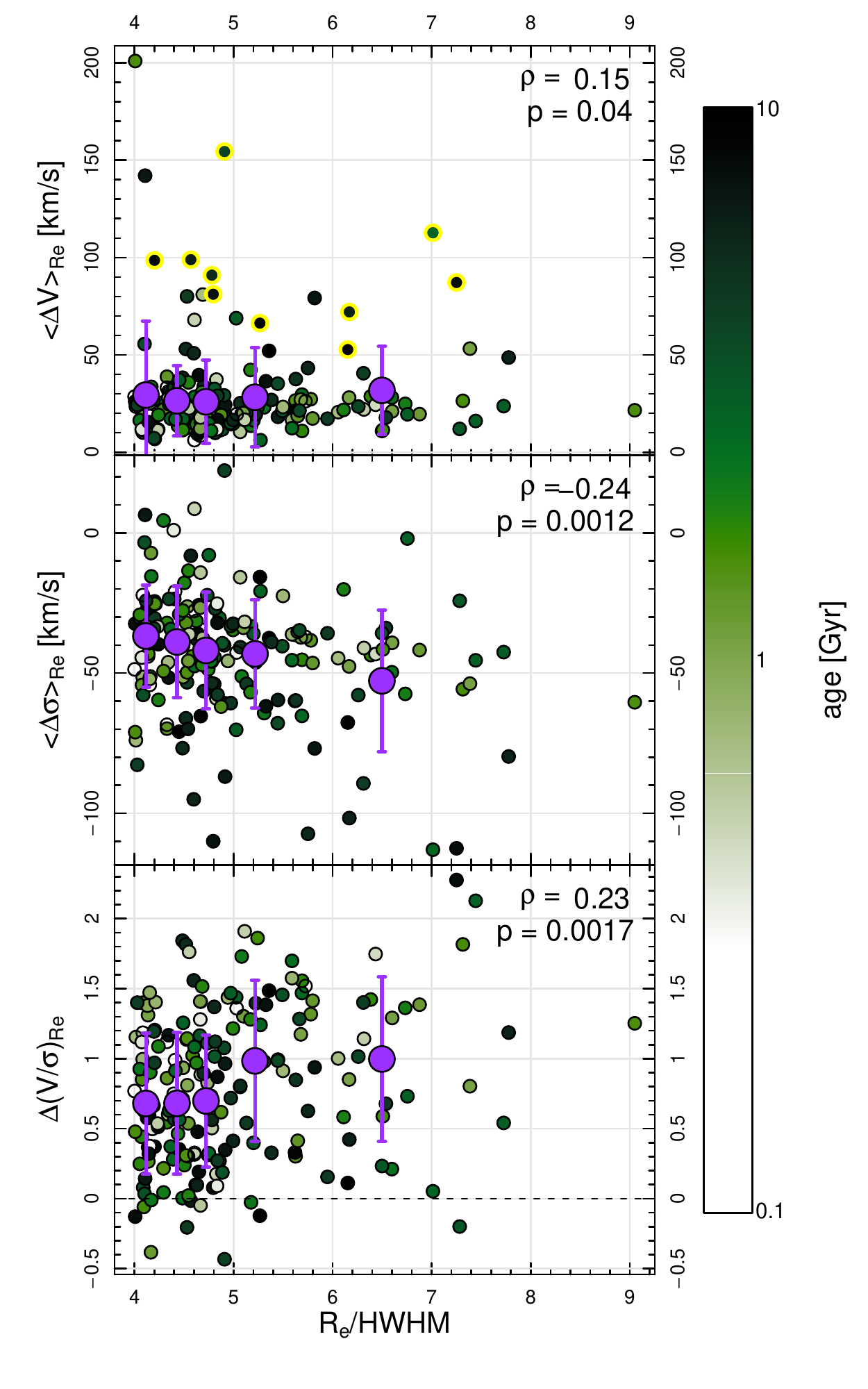}
\caption{Average local difference in velocity (top) and velocity dispersion (middle), and average global $\Delta(V/\sigma)$ (normalised to the stellar $(V/\sigma)$, bottom) between the ionised gas and stars as a function spatial resolution $R_e/ {\rm HWHM}$. Data points with yellow outlines in the top panel show systems with kinematically misaligned gas and stars. $R_e/ {\rm HWHM}$-binned mean values are shown as purple symbols with errorbars corresponding to the standard deviation of each bin.  Dashed line shows parity. Data points are colour coded by stellar age. The Spearman correlation coefficient $\rho$ and the corresponding $p$-value are stated in the top right of each panel. While these show trends (albeit only a weak on in the top panel), these go contrary to expectations if they were caused by beam smearing (see text).}\label{fig:hwhm_deltakin}
\end{figure}

First, we look for possible trends between the spatial resolution defined as the ratio of the size of the galaxy and the atmospheric seeing ($R_e/{\rm HWHM}$). In Fig. \ref{fig:hwhm_deltakin}, we find a negative trend between $\left<\Delta\sigma\right>_{R_e}$ and the spatial resolution ($R_e/ {\rm HWHM}$) and a weak trend with ($\Delta(V/\sigma)_{R_e}/(V/\sigma)_{\rm stars, R_e}$). Importantly, these trends go contrary to what would be expected should beam smearing be the driver i.e., the dynamical differences between gas and stars are more pronounced in galaxies with higher and thus more favourable $R_e/{\rm HWHM}$. We suggest that given our strict selection, these trends are instead dominated by real correlations between the physical size of galaxies ($R_e$) and dynamical differences between ionised gas and stars.

\begin{figure*}
\centering
\includegraphics[width=17cm]{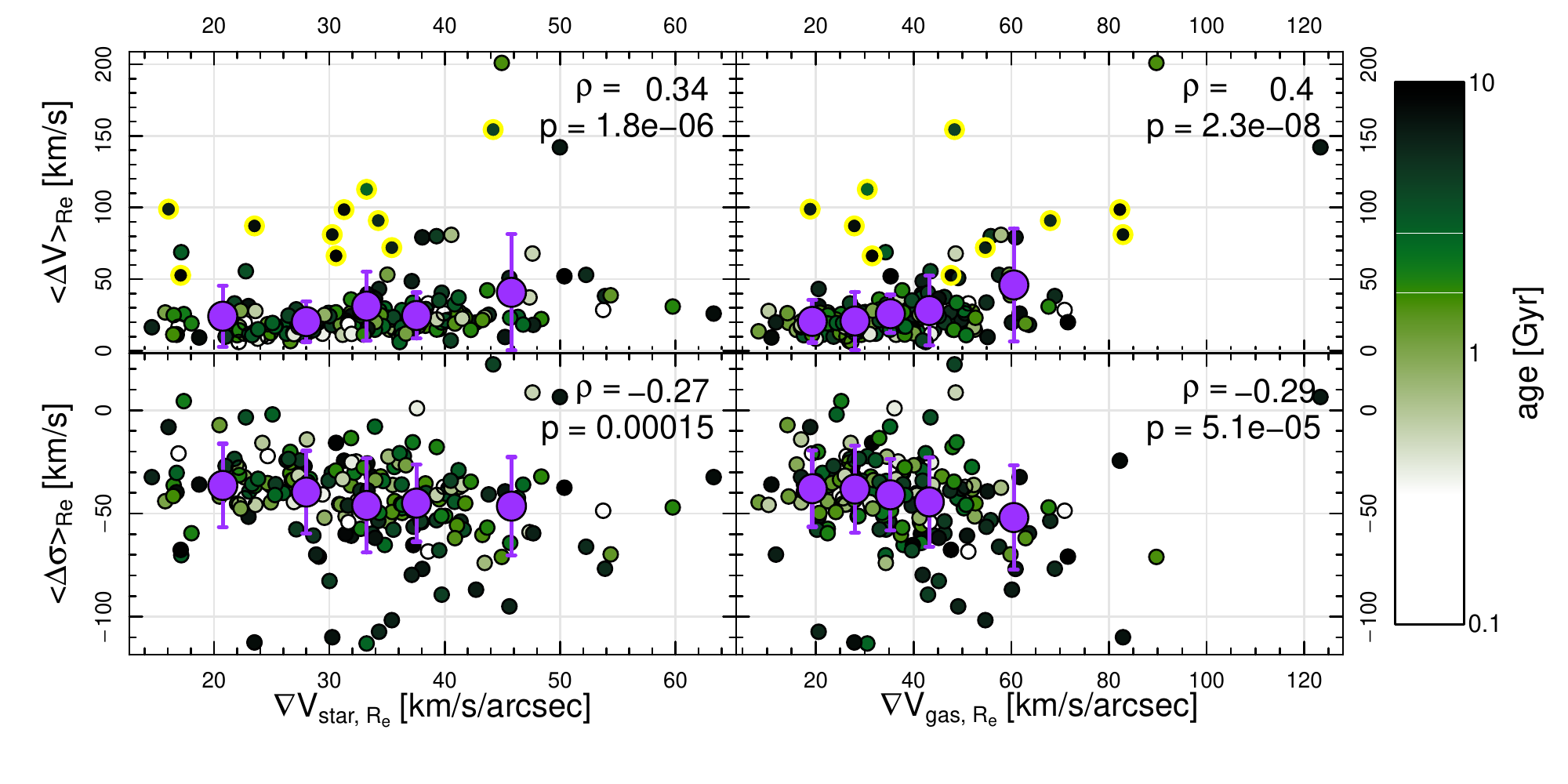}
\caption{Local velocity (top) and velocity dispersion (bottom) as a function of stellar (left) and gas velocity gradient (right). Data points with yellow outlines in the top panels show systems with kinematically misaligned gas and stars. Stellar (left) and gas (right) velocity gradient-binned mean values are shown as purple symbols with errorbars corresponding to the standard deviation of each bin. The Spearman correlation coefficient $\rho$ and the corresponding $p$-value are stated in the top right of each panel. The trend in $\left<\Delta V\right>_{R_e}$ with gas velocity gradient is more pronounced and statistically significant than that of $\left<\Delta \sigma\right>_{R_e}$. Trends with stellar velocity gradient are also less significant than those with gas velocity gradient. We argue these trends are at least partly due to real physical phenomena and cannot be entirely ascribed to beam smearing (see text).
}\label{fig:grad_deltakin}
\end{figure*}

Finally, following \citet{Varidel16} and \citet{Oh22}, we look for trends between the gas and stellar velocity gradients and our local and global kinematic measurements. Fig. \ref{fig:grad_deltakin} shows significant trends in $\left<\Delta V\right>_{R_e}$ with velocity gradients, which can be attributed to real differences between the stellar and ionised gas velocity gradients (see Fig. {\ref{fig:age_gradongrad}}). The correlation between $\left<\Delta \sigma\right>_{R_e}$ and velocity gradients is somewhat weaker, and less pronounced than those found in \citet{Oh22} for the dispersion ratio. This apparent discrepancy can be attributed to our markedly different choice of parameter definition and our more stringent sample selection. Given the results of previous tests in this Appendix, we instead attribute the bulk of the (admittedly sometimes weak) trends between dynamical differences and velocity gradients to real physical differences in the dynamics of the gas and stars already discussed in this work.

\bsp	
\label{lastpage}
\end{document}